\documentclass[]{kluconf}

\usepackage{graphicx}
\usepackage[latin1]{inputenc}
\usepackage{epsfig}
\usepackage{amsmath}
\usepackage{amsfonts}
\usepackage{amscd}
\usepackage{multirow}
\title{Bayesian Network Tomography and Inference}
\titlemark{Pluch Ph., Wakounig S.}

\author{Philipp Pluch and Samo Wakounig}
\authoraddress{
University of Klagenfurt \\ Department of Statistics \\
philipp.pluch@uni-klu.ac.at\\
\and \\
ARC Seibersdorf research GmbH\\ Department of Quantum
Cryptography\\
swakouni@edu.uni-klu.ac.at
}

\begin{document}

\maketitle
\section{Abstract}
%
%
%
%
%
%
The aim of this technical report is to give a short overview
of known techniques for network tomography (introduced in the
paper of Vardi (1996)), extended by a Bayesian approach originating 
Tebaldi and West (1998). Since the studies of A.K. Erlang
(1878-1929) on telephone networks in the last millennium, lots of
needs are seen in todays applications of networks and network
tomography, so for instance networks are a critical component of
the information structure supporting finance, commerce and even
civil and national defence. An attack on a network can be
performed as an intrusion in the network or as sending a lot of
fault information and disturbing the network flow. Such attacks
can be detected by modelling the traffic flows in a network, by
counting the source destination packets and even by measuring
counts over time and by drawing a comparison with this 'time
series' for instance.

\section{Introduction}

\begin{figure}[ht]
\centering \epsfig{file=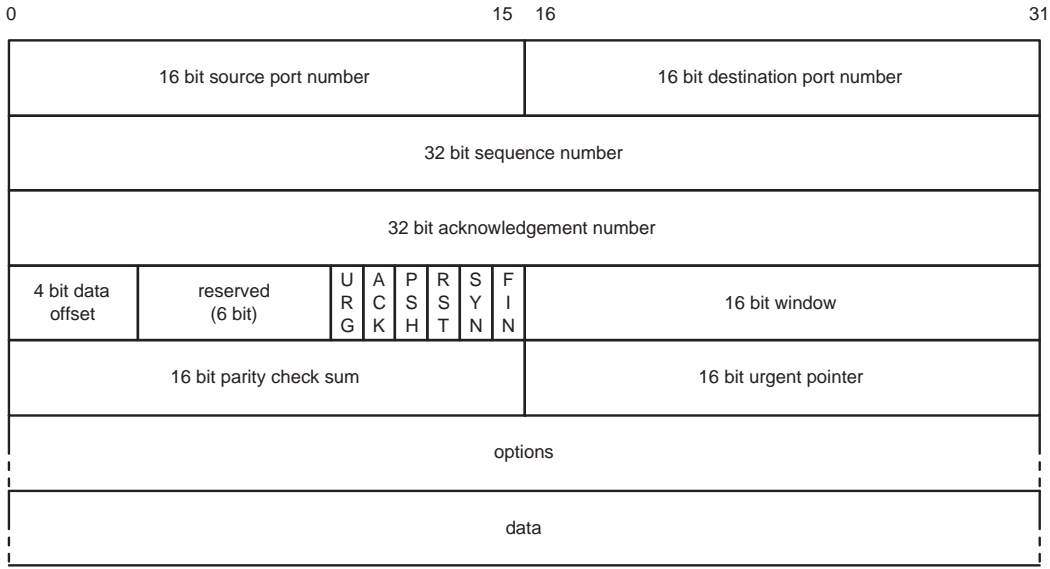, scale=0.9} \caption{The TCP
segment \label{TCPHead}}
\end{figure}

In order to know, find and understand the typical denial of service attack, it
is necessary to understand the 
principle of protocols transmitted over a network. As an example
we can look at the TCP protocol. The TCP-header is illustrated in
figure \ref{TCPHead}.

\begin{figure}[b]
\centering \epsfig{file=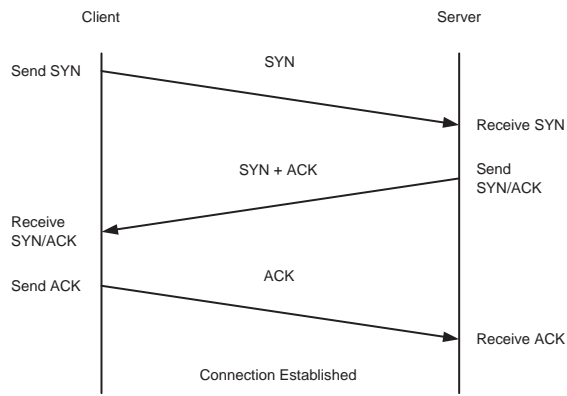, scale=0.75}
\caption{Three-Way-Handshake \label{3WShake}}
\end{figure}

\begin{figure}[t]
\centering \epsfig{file=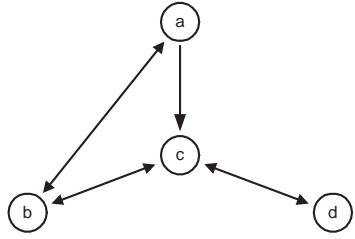, scale=0.8} \caption{Example
of a directed network with four nodes \label{nettomo1}}
\end{figure}

One of the main features of TCP is the concept of ports. Each
session to or from an application is assigned a source port and a
destination port. A source destination port pairing is used to
disambiguate multiple ongoing sessions between machines. TCP also
implements a two way connection scheme based on the usage of flags
in the header. A common interaction on a network will be, that the
client sends a packet with a synchronize flag set indication a
communication. The destination then opens a port for the
communication based on that request.
The server 
is responding with
both a synchronize and an acknowledgement flag and then finally
the client responds with an acknowledgement flag. This is the
three way handshake (see figure \ref{3WShake}),
which sets up the connection and allows a two way communication
(description in a very simply way). One idea of an attack is to
flood a computer with bogus requests or to cause it to devote
resources to the attack at the expense of the legitimate user of
the system. The attacker is just sending packets that request for
a communication but never completes the three way handshake.
Another attack is to send packets via the network that are full of
errors so that the victim computer is forced to spend time with
these errors. This results in a number of reset (or other) packets
with no obvious session.\\
%
%
%
%
%
%
We are able to compute detection probabilities in the following
way (see Marchette (2005) and Moore et al. (2001)). IP addresses
are unique 32-bit numeric address of every host and router on the
Internet. "`Spoofing"' denotes the changing of the source address
to a nonexistent address. We simply collect the packets with no
obvious session. Assume the spoofed IP addresses are generated
randomly, uniformly and independently on all $2^{32}$ addresses.
We assume that
$d$ packets are sent in an attack on a victim in a network. If we
monitor all packets to IP addresses, the probability of detecting
an attack is given by
\begin{eqnarray*}
  P(\textrm{'detect an attack'})=1-\left(1-\frac{w}{2^{32}}\right)^{d}
\end{eqnarray*}
with expected number of disturbing packets is given by
\begin{eqnarray*}
  \frac{wd}{2^{32}},
\end{eqnarray*}
where $w$ denotes the number of monitored IP addresses.
%
%
%
%
To infer how many packets were originally sent we need to estimate
the severity of an attack. Under the assumption of independence probability of defining $j$ packets
as attacking packets is given by 
\begin{eqnarray*}
  P(j \textrm{ 'packets'})= {d \choose
    j}\left(\frac{w}{2^{32}}\right)^{j}\left(1-\frac{w}{2^{32}}\right)^{d-j}
\end{eqnarray*}
and the maximum likelihood estimate for $d$ is given by
\begin{eqnarray*}
  \hat{d}=\left\lfloor \frac{j \, 2^{32}}{w} \right\rfloor.
\end{eqnarray*}
So if we see $j$ packets, we can estimate the number of such
attacks. From the literature it is known that the assumption of independence the number of
attack packets between two detected packets, is given by
\begin{eqnarray*}
  \sum_{s=1}^{w}s\left(1-\frac{w}{N}\right)^{s-1}\frac{w}{N},
\end{eqnarray*}
where $N$ is the number of IP addresses used for randomly simulating
those IP addresses that are used by an attacker, $w$ is the number
of monitored IP addresses. \\
Another more sophisticated approach will be in monitoring and
modelling the behaviour of the packets in the flow through the
network. Looking at the traffic we want to estimate the network
flow intensity.
The aim will be to estimate the traffic intensities by two ways.
First we are able to measure source destination (directed) pairs
of nodes and then perform repeated measurements on the nodes to
count packets (for example phone calls in routing, emails and so
on) transmitted over a communication network.
One main assumption in our mathematical model is, that we deal
with a strongly connected network, which means, that there always
exists a directed path between any two nodes. \\
When we study the architecture of networks, we distinguish between
two main groups of networks -- those that are deterministic (fixed
routing) networks and those that are random (Markovian routing)
networks. In the first group we deal with directed paths between
the nodes, that are fixed and known for each communication. For
the second group the travelling of information (sending of
packets) is determined by a fixed known Markov chain. It can be
easily seen
that random routing is a special case of fixed routing. \\
A source destination pair (short SD) transports information from
the source to the destination over a direct connected path in the
network. We introduce $c$ as the number of SD pairs, which can be
calculated from the number of nodes $n$ by
\begin{eqnarray*}
  c=(n-1)n.
\end{eqnarray*}
The number of transmitted information of a SD pair $j$ at
measurement period $k$ is given by $X_{j}^{(k)}$, which, like in classical
"teletraffic theory" is assumed to follow a Poisson distribution with parameter
$\lambda_{j}$, i.e.
\begin{eqnarray*}
  X_{j}^{(k)}\sim \textrm{Po}(\lambda_{j}).
\end{eqnarray*}
We can formulate the SD transmission vector at period $k$ by
$\mathbf{X}^{(k)}=(X_{1}^{(k)},...,X_{c}^{(k)})^{t}$. For the
modelling of the problem we need to introduce the $r\times c$
routing matrix $\mathbf{A}$ for a deterministic network as a
$(0,1)$-matrix given by
\begin{eqnarray*}
  \mathbf{A}=(a_{ij}).
\end{eqnarray*}
We get $a_{ij}=1$ if the link $i$ belongs to the directed path of
the SD pair and $a_{ij}=0$ if the link $i$ does not belong to
the directed path of the SD pair.\\
\\
Vardi (1996) gives several example networks, for instance
such network is given in figure \ref{nettomo1}.
It is a four-node directed network and consists of
\[c = (n-1)\cdot n =3\cdot4=12\]
SD pairs and seven directed links.\\
%
For better reading the routing matrix $\mathbf{A}$ is given for better reading in the
table \ref{AMatr} with $Y_i$ and $X_j$ describing the structure displayed in table \ref{struct}.\\
\begin{table}[h]\centering\begin{tabular}{|c|c|c|c|c|c|c|c|c|c|c|c|c|}\hline
$\mathbf{A}$ & $X_1$ & $X_2$ & $X_3$ & $X_4$ & $X_5$ & $X_6$ & $X_7$ & $X_8$ & $X_9$ & $X_{10}$ & $X_{11}$ & $X_{12}$\\
\hline
 $Y_1$  & 1 & 0 & 0 & 0 & 0 & 0 & 0 & 0 & 0 & 0 &
 0 & 0\\
 $Y_2$  & 0 & 1 & 1 & 0 & 0 & 1 & 0 & 0 & 0 & 0 &
 0 & 0\\
 $Y_3$  & 0 & 0 & 0 & 1 & 0 & 1 & 1 & 0 & 0 & 1 &
 0 & 0\\
 $Y_4$  & 0 & 0 & 0 & 0 & 1 & 0 & 0 & 0 & 0 & 0 &
 0 & 0\\
 $Y_5$  & 0 & 0 & 0 & 0 & 0 & 0 & 1 & 1 & 0 & 1 &
 1 & 0\\
 $Y_6$  & 0 & 0 & 1 & 0 & 0 & 1 & 0 & 0 & 1 & 0 &
 0 & 0\\
 $Y_7$  & 0 & 0 & 0 & 0 & 0 & 0 & 0 & 0 & 0 & 1 &
 1 & 1\\ \hline
\end{tabular}
\label{AMatr}\caption{Routing matrix $\mathbf{A}$}
\end{table}
\begin{table}[h]\centering
\begin{tabular}{clcl}
\par & $Y_1: \; a \rightarrow b$ & \par & $X_1: \; a \rightarrow b$\\
\par & $Y_2: \; a \rightarrow c$ & \par & $X_2: \; a \rightarrow c$\\
\par & $Y_3: \; b \rightarrow a$ & \par & $X_3: \; a \rightarrow c \rightarrow d$\\
\par & $Y_4: \; b \rightarrow c$ & \par & $X_4: \; b \rightarrow a$\\
\par & $Y_5: \; c \rightarrow b$ & \par & $X_5: \; b \rightarrow c$\\
\par & $Y_6: \; c \rightarrow d$ & \par & $X_6: \; b \rightarrow a \rightarrow c \rightarrow d$\\
\par & $Y_7: \; d \rightarrow c$ & \par & $X_7: \; c \rightarrow b \rightarrow a$\\
\par &  & \par & $X_8: \; c \rightarrow b$\\
\par &  & \par & $X_9: \; c \rightarrow d$\\
\par &  & \par & $X_{10}: \; d \rightarrow c \rightarrow b \rightarrow a$\\
\par &  & \par & $X_{11}: \; d \rightarrow c \rightarrow b$\\
\par &  & \par & $X_{12}: \; d \rightarrow c$
\end{tabular}
\label{struct}\caption{Structure represented in $\mathbf{A}$}
\end{table}
%
%
%
%
%
%
\\
\\
The measured data on all links of the network is given by
$\mathbf{Y}^{(k)}=(Y_{1}^{(k)},...,Y_{r}^{(k)})$, where $r$
denotes all directed links in the network with the property
$r=O(n)$ and $c>r$. The formulation of the network model is given
by
\begin{equation}
\mathbf{Y}=\mathbf{A X}\label{Grundform}
\end{equation}
and if we consider measurement periods $k$, we rewrite this as
\begin{eqnarray*}
\mathbf{Y}^{(k)}=\mathbf{A}\mathbf{X}^{(k)}.
\end{eqnarray*}
The goal is to estimate
$\boldsymbol{\lambda}=(\lambda_{1},...,\lambda_{c})$ from
$\mathbf{Y}^{(1)},...,\mathbf{Y}^{(k)}$. The following questions turn
up:
\renewcommand\labelitemi {\normalfont\bfseries\textendash}
\begin{itemize}
  \item Are the parameters identifiable?
  \item Are the estimates consistent?
\end{itemize}
\renewcommand\labelitemi {\textbullet}
The model we deal with is a linear one, but we cannot use a
linear regression nor a random effect model because we deal with a
$(0,1)$-matrix $\mathbf{A}$, nonnegativity constraints on the
parameters and the Poisson assumption for the number of
transmitted messages. \\
The identifiability of the parameter vector $\boldsymbol{\lambda}$
can be easily verified by the following lemma (see Vardi (1996)).\\
\\
\emph{Lemma}
  If the columns of the routing matrix $\mathbf{A}$ are all distinct and each column
  has at least one non-zero entry $a_{ij}$, then $\boldsymbol{\lambda}$ is
  identifiable.\\
\\
The proof follows the principle of induction and we refer to Vardi
(1996). If we find a zero column in the routing matrix
$\mathbf{A}$, we can conclude that the corresponding SD pair is
not connected by a path, and if there is a zero row, then the
corresponding link is not a part of the network. This
observations leads us to the following assertion:\\ \\
\emph{Lemma}
  If $c>2^{r}-1,$ then some rates $\lambda_{1},...,\lambda_{c}$ cannot be
  estimated separately.

\section{Parameter Estimation}

With this model setting we can apply classical maximum likelihood
estimation (MLE), also iterative expectation maximisation (EM)
algorithms are also proposed in the literature.  Using 
maximum likelihood estimation we can expect problems due to the
nonlinear constraints. The structure of the log-likelihood function,
which is to maximised, is hard to evaluate. The likelihood equations read
\begin{eqnarray*}
  \frac{\partial l}{\partial \lambda_{i}}=0 \textrm{    for
  }i=1,...,c,
\end{eqnarray*}
where $l$ denotes the logarithm of the likelihood function $L$.
In vector notation this can be expressed as
\begin{eqnarray*}
  \frac{1}{K}\sum_{k=1}^{K}E_{\boldsymbol{\lambda}}[\mathbf{X}^{(k)}|\mathbf{Y}^{(k)}=\mathbf{A}\mathbf{X}^{(k)}] -\boldsymbol{\lambda}=0
\end{eqnarray*}
$\mathbf{X}^{(k)}$ are the complete (unobserved) data and
$\mathbf{Y}^{(k)}$ are the incomplete (observed) data. A
formulation of the EM algorithm under the assumption of independence of the
$k$ components is given by 
\begin{eqnarray*}
  \boldsymbol{\lambda}^{(n+1)}=\frac{1}{K}\sum_{k=1}^{K}E[\mathbf{X}^{(k)}|\mathbf{Y}^{(k)},\boldsymbol{\lambda}^{(n)}].
\end{eqnarray*}
A problem which we mark out here is, that the above given summands
are hard to calculate, since the solutions are located in the
integer range. For finding a maximum it is necessary, that the
log-likelihood is concave. By evaluating the Hessian matrix
$\mathbf{H}$ given by
\begin{eqnarray*}
  \mathbf{H}=(\frac{\partial^{2}l}{\partial \lambda_{i} \partial
  \lambda_{j}}),
\end{eqnarray*}
we see that this matrix is not necessarily negative semidefinite, so $l$ is not necessarily concave (see
Vardi (1996)). A resolution of this problem is given by the following proposition \\
\\
\emph{Proposition 1} If $\boldsymbol{\lambda}^{\ast}$ is an interior
point, then for large $K$, $l$ is concave in the neighbourhood of
$\boldsymbol{\lambda}^{\ast}$.
\\
\\
Other estimation methods instead of MLE are based on normal
approximations. Under the assumption of normality:
\begin{eqnarray*}
  \mathbf{X}\sim N(\boldsymbol{\lambda}, \mathbf{\Lambda}),
\end{eqnarray*}
where $\mathbf{\Lambda} = \textrm{diag}(\boldsymbol{\lambda})$ is
a $c\times c$ matrix, the joint distribution of
$(\mathbf{X}^{t},\mathbf{Y}^{t})^{t}$ assuming
$\mathbf{Y}=\mathbf{AX}$ to hold is given by
\begin{eqnarray*}
  \left ( \begin{array}{c}\mathbf{X}\\ \mathbf{Y}\end{array}\right) =
  N_{c+r}\left(\left(\begin{array}{c}\boldsymbol{\lambda}\\ \mathbf{A}\boldsymbol{\lambda}
    \end{array}\right), \left( \begin{array}{cc}\boldsymbol{\lambda} &
      \mathbf{\Lambda}\mathbf{A}^{t} \\ \mathbf{A \Lambda} & \mathbf{A
        \Lambda} \mathbf{A}^{t}\end{array}\right)\right).
\end{eqnarray*}
The conditional distribution of $\mathbf{X}$ given $\mathbf{Y}$ is given by
\begin{eqnarray*}
  \mathbf{X}|\mathbf{Y}\sim
  N_{c}(\boldsymbol{\lambda}+\mathbf{\Lambda}\mathbf{A}^{t}(\mathbf{A \Lambda
    A})^{-1}(\mathbf{Y} -\mathbf{A}\boldsymbol{\lambda}), \mathbf{\Lambda}-\mathbf{
    A}^{t}(\mathbf{A\Lambda}\mathbf{A}^{t})^{-1}\mathbf{A\Lambda})
\end{eqnarray*}
so we are able to approximate
\begin{eqnarray*}
  E[\mathbf{X}|\mathbf{Y},\boldsymbol{\lambda}] \approx \boldsymbol{\lambda}+\mathbf{\Lambda}
  \mathbf{A}^{t}(\mathbf{A \Lambda A}^{t})^{-1}(\mathbf{Y}-\mathbf{A \Lambda})
\end{eqnarray*}
and get the following iteration formulae
\begin{eqnarray*}
  \boldsymbol{\lambda}^{(n+1)}=\frac{1}{K}\sum_{k=1}^{K}[\boldsymbol{\lambda}^{(n)}+
  \mathbf{\Lambda}^{(n)}\mathbf{A}^{t}(\mathbf{A
    \Lambda}^{(n)}\mathbf{A}^{t})^{-1}
    (\mathbf{Y}^{(k)}-\mathbf{A}\boldsymbol{\lambda}^{(n)})].
\end{eqnarray*}
At this point we should mention, that a priori all
$\lambda_{i}>>0$. Here we have to expect nonnegligible approximation
errors. Because of the matrix inversion some of the
$\boldsymbol{\lambda}^{(i+1)}$ can be negative. Another approach
that has been proposed in the literature is to assume the sum of the $\mathbf{Y}^{(k)}$
to be normally distributed,
\begin{eqnarray*}
  \bar{\mathbf{Y}}=\frac{1}{K}\sum_{k=1}^{K}\mathbf{Y}^{(k)} \sim
  N_{r}(\mathbf{A}\boldsymbol{\lambda}, K^{-1}\mathbf{A\Lambda
  A}^{t}).
\end{eqnarray*}
The log-likelihood of $\bar{\mathbf{Y}}$ is given by
\begin{eqnarray*}
  l(\boldsymbol{\lambda})=-\log |\mathbf{A\Lambda A}^{t}|-K(\bar{\mathbf{Y}}
  -\mathbf{A}\boldsymbol{\lambda})^{t}(\mathbf{A \Lambda
    A}^{t})^{-1}(\bar{\mathbf{Y}}-\mathbf{A}\boldsymbol{\lambda}) \rightarrow
  \max_{\lambda_{i}\geq 0}.
\end{eqnarray*}
Another approach to the estimation problem is based on sample moments
where $\bar{\mathbf{Y}}$ is completely determined by the mean
vector $\mathbf{A}\boldsymbol{\lambda}$ and the covariance matrix
$\mathbf{A\Lambda A}^{t}$ of $\mathbf{Y}$ and under the usage of
the first and second moment we get
\begin{itemize}
  \item $\widehat{E(\mathbf{Y})}=\bar{\mathbf{Y}}=\mathbf{A}\boldsymbol{\lambda}$
  \item $\textrm{Cov}(Y_{i},Y_{h})=\frac{1}{K}\sum_{k}Y_{i}^{(k)}Y_{h}^{(k)}
    -\bar{Y_{i}}\bar{Y_{h}}=\sum_{l=1}^{c}a_{il}a_{hl}\lambda_{l}$  for $1\leq
    i \leq h \leq r$
\end{itemize}
where the first moment equation is independent of the Poisson
assumption and the second moment equation strongly depends on the
Poisson model.
\section{Prior Models for Network Tomography}
In this section we will investigate the problem of computing and summarizing
the joint posterior distribution of $p(\mathbf{X}|\mathbf{Y})$ for
all observed messages of SD pairs given the observed link counts
$\mathbf{Y}$. For the posterior distribution we need a model for
the prior distribution $p(\mathbf{X})$ to be tied together with
$\mathbf{Y}=\mathbf{AX}$. Under the assumption, that the $X_j$ are
independently Poisson distributed over the routs $j$, the prior
specification is completed by a prior of $\mathbf{\Lambda}$. The
joint distribution of the model is then given by
\begin{eqnarray*}
  p(\mathbf{X},\mathbf{\Lambda})=p(\mathbf{\Lambda})\prod_{j=1}^{c}
  \lambda_{j}^{X_{j}}e^{\frac{-\lambda_{j}}{X_{j}!}}.
\end{eqnarray*}
We are interested in estimation of $\mathbf{X}$ since we can
infer $\mathbf{X}$ and $\mathbf{\Lambda}$ from the joint
distribution. On a more advanced modelling standard we can also
use hierarchical modeling for the parameters $\lambda_{i}$. It is
common in literature to model such hyperparameters by a normal
distribution $N(\mu, \sigma)$. Posterior computations are
difficult to evaluate analytically. For example, it is unrealistic
to evaluate them for large networks. For this purpose we introduce
some iterative MCMC simulation algorithms. As an example consider
Gibbs sampling, which iteratively resamples from the conditional
posterior for elements of the $\mathbf{X}$ and $\mathbf{\Lambda}$
variables. Under the usage of
\begin{eqnarray*}
  p(\mathbf{\Lambda}|\mathbf{X,Y})=p(\mathbf{\Lambda}|\mathbf{X})=
  \prod_{j=1}^{c}p(\lambda_{j}|X_{j}),
\end{eqnarray*}
whose components have the form of the prior density
$p(\lambda_{j})$ multiplied by a gamma function arising in the
Poisson based likelihood function. It is possible to simulate new
$\mathbf{\Lambda}$ values as a set of independent drawing from the
univariate posterior density. If the prior densities are gamma
densities or a mixture of gamma densities, these drawings are
trivially made from the corresponding gamma or mixed gamma
posterior densities. Otherwise, as proposed in literature, we have
to use the rejection method or embedded Metropolis Hasting steps
in the MCMC scheme in the standard
Metropolis - Gibbs framework. \\
The theoretical structure of a general network model leads to the
following theoretical result for computing samples from the
conditional posteriors. Let $\mathbf{\Lambda}$ be fixed and focus
on the conditional posterior $p(\mathbf{X}|\mathbf{\Lambda,Y})$,
then we can use the following theorem due to Tebaldi and
West (1998).
\\ \\
\emph{Proposition 2} In the network model $\mathbf{Y}=\mathbf{AX}$ and
under the assumption that $\mathbf{A}$ has full rank $r$, we can
reorder the columns of $\mathbf{A}$ such, that the revised routing
matrix has the form
\begin{eqnarray*}
  \mathbf{A}=[\mathbf{A}_{1},\mathbf{A}_{2}]
\end{eqnarray*}
where $\mathbf{A}_{1}$ is a nonsingular $r\times r$ matrix. By
similar reordering the elements of the vector $\mathbf{X}$ and
partitioning as $\mathbf{X}^{t}=(\mathbf{X}_{1}^{t},
\mathbf{X}_{2}^{t})$, it follows that
\begin{eqnarray*}
  \mathbf{X}_{1}=\mathbf{A}_{1}^{-1}(\mathbf{Y}-\mathbf{A}_{2}\mathbf{X}_{2}).
\end{eqnarray*}
This result easily follows from the fact that
$\mathbf{Y}=\mathbf{AX}=\mathbf{A}_{1}\mathbf{X}_{1} +
\mathbf{A}_{2}\mathbf{X}_{2}$.\\
\\
The full rank assumption is satisfied by all networks in real
world. Otherwise there is a redundancy in specification, and one
or more rows of $\mathbf{A}$ can be deleted to get linear
independent rows. The result in the last proposition implies that,
given $\mathbf{Y}$ and the assumed values of the $(c-r)$ route
counts in $\mathbf{X}_{2}$, we are able to compute directly the
remaining $r$ route flows simply based on the algebraic structure
of the routing matrix. For the reordering of the matrix
$\mathbf{A}$ we can use the QR decomposition of arbitrary full
rank matrices. After the QR decomposition we get an $r \times r$
orthogonal matrix $\mathbf{Q}$ and an $r\times c$ upper triangular
matrix $\mathbf{R}$, the first $r$ columns of which correspond to $r$
linear independent columns of $\mathbf{A}$. These are identified
by a permutation of column indices. \\
With this knowledge we can deduce, that the conditional
distribution $p(\mathbf{X}|\mathbf{\Lambda,Y})$ lies in the $c-r$
dimensional subspace defined by the partition
$\mathbf{A}=[\mathbf{A}_{1},\mathbf{A}_{2}]$. After the partitioning
of the routing matrix $\mathbf{A}$ the posterior has the form

\begin{eqnarray*}
  p(\mathbf{X_1|X_2,\Lambda,Y})p(\mathbf{X_2|\Lambda,Y}),
\end{eqnarray*}

where $p(\mathbf{X_1|X_2,\Lambda,Y})$ is degenerated at
$\mathbf{X_1 =A_1^{-1}(Y-A_2 X_2)}$ with $\mathbf{X}_{2}=(X_{r+1},
..,X_{c})^{t}$ and $\mathbf{X}_{1}=(X_{1},...,X_{r})^{t}$ defined
as above. The conditional is given by
\[p(\mathbf{X_2|\Lambda,Y})\propto \prod_{a=1}^c \frac
{\lambda_a^{X_a}}{X_a!}\] with the support defined by $X_a\geq 0$
for all $a=1,\ldots, c$. It is the product of independent Poisson
priors for the $X_{i}$ constrained by the model and the
reordering. The utility of this expression is in delivering the
set of complete conditional posteriors for elements of the
$\mathbf{X_2}$ vector to form a part of the iterative simulation
approach to posterior analysis. Let´s now consider each element
$X_i$ of $\mathbf{X_2}, (i=r+1,\ldots,c)$, and write
$\mathbf{X_{2,-i}}$ for the remaining elements. The conditional
distribution $p(X_i|\mathbf{X_{2,-i},\Lambda,Y})$ is given by
\[p(X_i|\mathbf{X_{2,-i},\Lambda,Y}) \propto
\frac{\lambda_i^{X_i}}{X_i!}\prod_{a=1}^r
\frac{\lambda_a^{X_a}}{X_a!},\] over the support of the expression
above. The linear constraints on $X_i, i\in \{r+1,\ldots,c\}$, are
of the form $X_i \geq d_i$ and $X_i \leq e_i,$ where the values
$d_i$ and $e_i$ are functions of the conditioning values of
$\mathbf{X_{2,-i}}$ and $\mathbf{Y}$. Together with $X_i \geq 0$
we obtain at most a set of $r+1$ constraints on $X_i$. It is
computationally very burdensome to evaluate directly these
constraints and identify their intersection. So we can make direct
simulations.
\\
For the simulation of the full posterior $p(\mathbf{X,\Lambda|Y})$
we need now fixed starting values of the route counts $\mathbf{X}$.
We can apply the following algorithm according to Tebaldi and West
(1998):
\\
\\
\emph{Algorithm}

\begin{enumerate}

\item Draw sampled values of the rates $\mathbf{\Lambda}$ from $c$
conditionally independent posteriors
$p(\lambda_a|X_a),$\label{one}

\item conditioning on these values of $\mathbf{\Lambda}$ simulate
a new $\mathbf{X}$ vector by sequencing through $i=r+1,\ldots,c$,
and at each step sample a new $X_i$ with the conditioning elements
from the $\mathbf{X_{2,-i}}$ set at their most recent sampled
values,\label{two}

\item iterate.\label{three}

\end{enumerate}

This is a known standard Gibbs sampling setup. Scalar elements of
both $\mathbf{\Lambda}$ and $\mathbf{X}$ are resampled from the
relevant distribution conditional on most recently simulated
values of all other uncertain quantities. In step 2 we require
evaluation of the support which is best done by a simulation
method such as embedded Metropolis-Hastings steps. We note that
from $\mathbf{Y=AX}$ it is possible to identify bounds on each
$X_i$. A suitable range for the proposal distribution can be
computed from that.
\section{Usage of Bayes Factors for Modelling of Network Traffic}
If there is a sequence of packets transmitted over a network, we
can evaluate a statistical profile of that sequence based on the
information of the header and compare this to similar sequences in
the past. This historical behaviour can be saved in a stochastic
matrix where each element of the matrix is given by
\begin{eqnarray*}
  p_{jku}=P(\textrm{'SD'}=k|\textrm{'SD before'}=j, \textrm{'IP of sender'} =u)
\end{eqnarray*}
Since we know the header in the packet, we are able to model the
behaviour of the sender over time. We can base an analysis on these
matrices. DuMouchel (1999) has made a similar approach for modelling of the behaviour of commands in a shell. Since
we use some kind of categorical data, we are able to use the
multinomial distribution as proposed in the literature. For
Bayesian inference we can use the Dirichlet (prior) distribution, which is the
natural
conjugate distribution to the multinomial distribution. \\
Let $\mathbf{p}=(p_{1},...,p_{K})$ be a random vector which is
Dirichlet distributed with the density
\begin{eqnarray*}
  f(\mathbf{p})=\frac{\Gamma(\sum_{k}\alpha_{k})\prod_{k}(p_{k}^{\alpha_{k-
        1}})}{\Gamma(\alpha_{k})}
\end{eqnarray*}
with $\alpha_{i}>0$ for all $i$. The multinomial probability for a
count data vector $\mathbf{n} = (n_{1},...,n_{K})$ with
$\tilde{n}=\sum_{k}n_{k}$ is given by
\begin{eqnarray*}
  P(n|p)=\tilde{n}!\prod_{k}\frac{p_{k}^{n_{k}}}{n_{k}!}
\end{eqnarray*}
From above formulas we get the marginal distribution
\begin{eqnarray*}
  P(n)=\frac{\tilde{n}!}{\tilde{\alpha}(\tilde{\alpha}+1)\cdot\ldots\cdot(\tilde{\alpha}+\tilde{n}-1)}\prod_{k}\frac{\alpha_{k}(\alpha_{k}+1)\cdot\ldots\cdot(\alpha_{k}+n_{k}-1)}
  {n_{k}!}
\end{eqnarray*}
with $\tilde{\alpha}=\sum_{k}\alpha_{k}$. The  posterior
distribution is given by the following Dirichlet distribution
\begin{eqnarray*}
  P(p|n)=(\tilde{\alpha}+\tilde{n}-1)!\prod_{k}\frac{p_{k}^{n_{k}+\alpha_{k}-1}}
  {(\alpha_{k}+n_{k}-1)!}
\end{eqnarray*}
On the idea that one user $u$ in the network generates a sequence
of $T+1$ packets $C_{0},C_{1},...,C_{T}$ we can build the following
hypotheses for a test of sending packets that disturb the network
\begin{eqnarray*}
  &H_{0}:P(C_{t}=k|C_{t-1}=j)=p_{jku}\\
  &H_{1}:P(C_{t}=k|C_{t-1}=j)=Q_{k}
\end{eqnarray*}
where
\begin{eqnarray*}
  (Q_{1},...,Q_{k})\sim
  \textrm{Dirichlet}(\alpha_{01},...,\alpha_{0k}).
\end{eqnarray*}
We make the assumption that the null hypothesis $H_{0}$ says that a
legitimate user is generating packets out of the profiles of the
transition probabilities. The alternative hypothesis $H_{1}$ says, that
$T$ packets are sent through the network, are drawn randomly and
independently from a probability vector following a Dirichlet
distribution with given hyperparameters. These hyperparameters have
to be estimated. $H_{1}$ is more general than $H_{0}$ since we do
not know $Q$ in comparison with the fully specified
$\mathbf{P}_{u}=(p_{jku})$. $H_{0}$ is not nested in $H_{1}$. For
checking the practicability we suggest the usage of Bayes factors
$BF$ given by
\begin{eqnarray*}
  BF=\frac{P(C_{0},...,C_{T}|H_{1})}{P(C_{0},...,C_{T}|H_{0})}
\end{eqnarray*}
for inference. For large $BF$ we will prefer the alternative
hypotheses. Instead of BF often
\begin{eqnarray*}
  x=\log(BF)
\end{eqnarray*}
is used, which is called the "weight of evidence". We can see
that modelling the behaviour using the network with the network
tomography and combining it with the concepts of Bayes factors we
are able to implement a large apparatus for monitoring networks
and to draw a conclusion whether there is an attack on our
monitored network.

\section{Conclusions}

In this paper we gave an overview of the current literature in
network modelling and network tomography. We extended that field
by developing a method for monitoring networks with Bayes
factors for testing hypothesis testing whether there is an intruder in the
network, who performs several forms of attacks. This method can be
implemented with the usage of control charts which are common in
quality control for monitoring networks. Further work will be in
random routing networks -- for an overview of current applications
we refer to Vardi (1996). The methodologies developed here should
be also applicable for these kinds of networks. Results and
implementation will be given in further technical reports.

\section{References}

\begin{description}
  \item[Castro, R., Coates, M., Liang, G., Nowak, R., Yu,B.](2004) \emph{Network
      Tomography: Recent Developments.} Statistical Science 19
      (3): 499--517
  \item[DuMouchel, W.](1999) \emph{Computer Intrusion Detection Based on Bayes Factors for
      Comparing Command Transition Probabilities.} National
      Institute of Statistical Sciences (NISS), Technical Report
      Number 91
  \item[Marchette, D. J.](2001) \emph{Computer Intrusion Detection and Network
    Monitoring: A Statistical Viewpoint.} New York: Springer
  \item[Marchette, D. J.](2005) \emph{Passive Detection of Denial of Service
      Attacks on the Internet.} In: Statistical Methods in Computer
    Security. Dekker: New York
  \item[McCulloch, R.](1998) \emph{Bayesian Inference on Network Traffic Using
      Link Count Data: Comment.} Journal of the American Statistical
      Association 93 (442): 575
  \item[Moore, D., Voelker, G.M., Savage, S.](2001) \emph{Inferring Internet
      Denial-of-Service Activity.} USENIX Security Symposium'01.
      www.usenix.org/publications\\/library/proceedings/sec01/moore.html
  \item[Tanenbaum, A. S.] (1996) \emph{Computer Networks.} Upper Saddle River: Prentice Hall
  \item[Tebaldi, C., West, M.] (1998) \emph{Bayesian Inference on Network Traffic
      Using Link Count Data.} Journal of the American Statistical
      Association 93 (442): 557--573
  \item[Vardi, Y.](1998) \emph{Bayesian Inference on Network Traffic Using
      Link Count Data: Comment.} Journal of the American Statistical
      Association 93 (442): 573--574
  \item[Vardi, Y.](1996) \emph{Network tomography: Estimating
      source-destination traffic intensities from link data.}  Journal of the
    American Statistical Association 91 (433): 365--377
  \item[Willinger, W., Paxson, V.](1998) \emph{Where mathematics meets the
  internet.} Notices of the American Mathematical Society 45 (8):
  961--970
 \item[Wakounig, S.](2005) \emph{Einfache statistische Ansätze
 für Intrusion Detection Systeme.} MSc Thesis. University of
 Klagenfurt. Department of Applied Statistics.

\end{description}

\end{document}